\documentclass[anonymous=false, %
               format=acmsmall, %
               review=false, %
               screen=true, %
               nonacm=true]{acmart}

\usepackage[ruled]{algorithm2e}
\usepackage{tikz}
\usetikzlibrary{fit,positioning}

\usepackage{amsmath}
\usepackage{natbib}
\usepackage{xcolor}

\usepackage[utf8]{inputenc}

\usepackage[utf8]{inputenc}

\DeclareFixedFont{\ttb}{T1}{txtt}{bx}{n}{8} 
\DeclareFixedFont{\ttm}{T1}{txtt}{m}{n}{8}  

\usepackage{color}
\definecolor{deepblue}{rgb}{0,0,0.5}
\definecolor{deepred}{rgb}{0.6,0,0}
\definecolor{deepgreen}{rgb}{0,0.5,0}
\definecolor{deeporange}{rgb}{0.6,0.25,0}
\definecolor{verylightgray}{rgb}{0.97,0.97,0.97}

\usepackage{listings}

\newcommand\pythonstyle{\lstset{
language=Python,
basicstyle=\footnotesize\ttm,
keywordstyle=\ttb\color{deepblue},
commentstyle=\ttm\color{deepgreen},
stringstyle=\color{deeporange},
emphstyle=\ttb\color{deepred},    
emph={MyClass,__init__},          
otherkeywords={self,yield},       
frame=single,                     
showstringspaces=false, 
breaklines=true,
backgroundcolor=\color{verylightgray},
}}

\lstnewenvironment{python}[1][]
{
\pythonstyle
\lstset{#1}
}
{}

\newcommand\pythoninline[1]{{\pythonstyle\lstinline!#1!}}
\renewcommand{\v}[1]{\pythoninline{#1}}

\begin{document}

\title{\texttt{tfp.mcmc}: Modern Markov Chain Monte Carlo Tools Built for Modern Hardware}

\author{Junpeng Lao*, Christopher Suter*, Ian Langmore, Cyril Chimisov, Ashish Saxena, Pavel Sountsov, Dave Moore, Rif A. Saurous, Matthew D. Hoffman, and Joshua V.~Dillon}
\thanks{*\text{The first two authors contributed equally}}

\begin{abstract}
Markov chain Monte Carlo (MCMC) is widely regarded as one of the most important algorithms of the 20th century \citep{brooks2011handbook}. Its guarantees of asymptotic convergence, stability, and estimator-variance bounds \citep{brooks2011handbook, robert2013monte, neal1993probabilistic} using only unnormalized probability functions make it indispensable to probabilistic programming. In this paper, we introduce the TensorFlow Probability MCMC toolkit, and discuss some of the considerations that motivated its design.
\end{abstract}

\affiliation{\institution{Google Research}}
\email{tfprobability@tensorflow.org}
\date{January 2020}

\maketitle

\section{Introduction}

The TensorFlow Probability (TFP) MCMC library design flows mainly from three
principles. First, TFP MCMC takes advantage of both vectorized computation and
threading for single- and multiple-chain parallelism. Second, the framework is
agnostic to any particular probabilistic modeling framework: TFP MCMC requires
only a Python \v{callable} to compute the target log probability
(TLP)\footnote{Generally the function implementation is presumed to be using
TensorFlow \citep{abadi2016tensorflow}. Some, but not all, TFP MCMC algorithms
also rely on TF's automatic differentiation functionality.}. Third, we present
a user-level API of composable building blocks for constructing new MCMC
transition kernels and for use in higher-level algorithms.

The following example depicts our three contributions as a working example.
First, notice that the \v{target_log_prob} automatically {\em leverages
vectorized hardware}., since the input \v{x}, initialized as
\v{current_state=tf.zeros([100])} is a \v{tf.Tensor}. The result is 100 MCMC
chains run in parallel. Second, notice that there is otherwise {\em no
domain-specific language} for specifying the TLP. Third, notice that the {\em
transition kernel} mechanics are governed, in this case, by
\v{tfp.mcmc.HamiltonianMonteCarlo} which is used as a black box by the
\v{tfp.mcmc.sample_chain} {\em kernel driver}.


\begin{python}
def target_log_prob(x):
  return -0.5 * x ** 2
hmc_kernel = tfp.mcmc.HamiltonianMonteCarlo(
    target_log_prob_fn=target_log_prob,
    num_leapfrog_steps=3,
    step_size=1.5)
samples, traced_kernel_results = tfp.mcmc.sample_chain( 
    kernel=hmc_kernel,         
    num_results=100, 
    num_burnin_steps=300, 
    trace_fn=lambda _, pkr: pkr.accepted_results.step_size, 
    current_state=tf.zeros([100]))
\end{python}

\section{First Contribution: Pervasive Data Parallelism}
Multi-chain MCMC is intrinsically embarrassingly parallel---each chain is an
independent computation.\footnote{Some algorithms---like replica exchange
Monte Carlo, discussed below---share information between chains, but each
chain can still perform a single step completely independently.} Most
widely-used MCMC frameworks achieve chain parallelism by running one chain per
available processing unit (``task parallelism'') \citep{carpenter2017stan,
salvatier2016probabilistic, bingham2019pyro}. While convenient, naive task
parallelism fails to leverage ``single instruction, multiple data'' (SIMD)
instruction sets (``data parallelism''). SIMD capabilities are ubiquitously
present on both general-purpose CPUs (e.g., AVX) and specialized processors
(e.g., GPUs, TPUs). Since SIMD is in principle not mutually exclusive with
task parallelism, it presents an opportunity to increase multi-chain MCMC
parallelism by multiple orders of magnitude. For example, a 32 core/64 thread
CPU (the desktop soon-to-be largest \href{https://www.amd.com/en/ryzen}{AMD Ryzen}) could run as
many as 1024 parallel MCMC chains by leveraging 16 way SIMD parallelism via
AVX512 (at float32 resolution).

Despite the prospect of significant speedups, SIMD parallelism across diverse
hardware has not seen wide adoption by popular MCMC frameworks. In order to
leverage SIMD parallelism, the MCMC framework  and the user-provided model
must \emph{both} support vectorization. In TFP, we refer to vectorizable code
as having ``batch semantics'' or simply ``batching'' \citep{dillon2017}.

We believe SIMD chain parallelism has not been widely adopted in MCMC
frameworks because it requires multiple challenging pieces to fall into
place:
\begin{itemize}
    \item[-] The framework for specifying the model or TLP function must
    support batching.
    \item[-] The user-provided model or TLP function must be written to
    support batching (see code below for an example and counterexample).
    \item[-] The MCMC framework itself must at a minimum not \emph{impede}
    batching and in some cases (discussed below) must take significant pains
    to support it.
\end{itemize}
Moreover, the framework should handle batches indexed by arbitrary dimensions.

The following \emph{seemingly} identical TLP functions illustrate the subtlety
of proper batch semantics. While all are meant to produce independent Gaussian
samples, only the last two compute the correct answer when the input is
matrix-variate, i.e., with ``list of vectors'' semantics.

\begin{python}
def batch_hostile_target_log_prob(x):
  return -0.5 * tf.reduce_sum(x**2)           # Sums over batches too. Oops!
def batch_friendly_target_log_prob(x):
  return -0.5 * tf.reduce_sum(x**2, axis=-1)  # Sums over just the event. Yay!
def batch_easy_target_log_prob(x):
  return tfd.MultivariateNormalDiag(loc=tf.zeros(n_dims), scale=1.).log_prob(x)
\end{python}

As illustrated by this example, a primary objective of the overall TFP library
is to provide powerful and ergonomic tools for building complex probabilistic
models which leverage SIMD parallelism on diverse hardware ``out-of-the-box''
\citep{dillon2017}. This is in large part possible because TensorFlow itself
pervasively supports batching and execution on diverse hardware
\citep{abadi2016tensorflow}. By extending this ethos into probabilistic
programming, TFP provides users with a rich modeling language that hides many
of the challenges of vectorizing computation. We note, however, that the TFP
MCMC toolbox does not require that TLP functions be built from TFP
abstractions (e.g., Distributions, Bijectors). The library is ``self
agnostic,'' and a user may directly use TensorFlow if they prefer.

We now highlight several case studies internal to TFP MCMC's implementation.
These cases span from (1) simply needing to ``get out of the way'' of
batching, to (2) requiring batch parallelism for great profit, to (3) the
sometimes significant engineering challenges of maintaining vectorized
computation.

\subsection{Case Study: HMC}
Preserving vectorized computation for Hamiltonian Monte Carlo (HMC,
\citep{neal2011mcmc}) and other similarly ``simple'' Metropolis-Hastings
\citep{hastings1970monte} samplers is generally straightforward. In the case
of HMC specifically, one samples from a momentum distribution and uses an
iterative symplectic integrator with a fixed number steps to produce a
proposal. The proposal is accepted or rejected according to a
Metropolis-Hastings scheme. Each of these operations is trivially batchable.
We must be mindful to sample the momenta in accordance with the shapes of the
inputs---one independent momentum per state variable per chain---and to ensure
the user-provided step size is appropriately broadcast across the state space.

Throughout TFP MCMC, the TLP for multiple chains is evaluated as a batch. For
example, the HMC transition kernel may query the probability of a
vector-variate distribution with a matrix of values, i.e., a ``list of
vectors,'' wherein each row corresponds to a different chain state. Assuming
the supplied TLP function has ``batch semantics''---and the list length is
less than the processor's SIMD capacity---this evaluation costs roughly the
same as evaluating one chain.

Vectorization of control flow, such as the Metropolis-Hastings accept/reject
decision, can be subtle. For instance, TensorFlow has two ops for implementing
conditional logic: \v{tf.where} and \v{tf.cond}. The former is vectorized and
the latter is not. TFP MCMC generally avoids using \v{tf.cond}.

\subsection{Case Study: Replica Exchange Monte Carlo}
Replica Exchange MCMC (REMC) is a technique for increasing mixing rates of
multi-modal TLP functions. Also known as ``parallel tempering'', it is useful
when the modes of the TLP are separated by relatively large (with respect to
the chain variance), low-probability regions. REMC generates a single sample
by accepting or rejecting swaps among $K$ replicas, where replica $i$ has TLP
$p(x) / T_i$, $1 = T_1 < \cdots < T_K$.  In TFP, an \emph{additional} batch
dimension indexes the $K$ replicas, so that each replica may be a number of
independent chains, all sampling from $p(x)/T_i$.

The batch dimension indexing replicas is added by the REMC kernel.  By
assuming the user-supplied TLP and kernel driver have batch semantics, the TFP
MCMC REMC implementation need only replicate additional temperature-scaled
chains per each original chain.  This is essentially accomplished by the
following:

\begin{python}
def _make_replica_target_log_prob_fn(target_log_prob_fn, inverse_temperatures):
  def _replica_target_log_prob(*x):
    tlp = target_log_prob_fn(*x)
    # Pad shape on the right with 1's until its rank matches that of tlp.
    brodacastable = mcmc_util.left_justified_expand_dims_like(
        inverse_temperatures, tlp)
    # Scale tlp's by inverse_temperatures, taking care to match numeric types.
    return tf.cast(broadcastable, tlp.dtype) * tlp
  return _replica_target_log_prob
\end{python}

That is, temperature scaling across chains is achieved simply by broadcasting.
Similarly, we can easily implement other population-based samplers like
Differential Evolution MCMC \citep{ter2006markov} as information across chains
is readily available.

\subsection{Case Study: No U-Turn Sampler}
The No U-Turn Sampler (NUTS) \citep{hoffman2014no} has emerged as a popular
MCMC technique, in part for its more-“turnkey” nature (vs HMC). NUTS builds on
HMC by sampling from a Hamiltonian trajectory constructed by a recursive
doubling algorithm. The recursion is a pre-order tree traversal that
terminates when the trajectory makes a U-turn.\footnote{Another termination
criterion is ``divergence''---an absolute change in Hamiltonian energy above
some threshold.} The number of recursions is therefore dynamic, varying based
on the starting position and momentum, and so some chains may incur more
recursions than others. Achieving data parallelism for NUTS is made
challenging by the recursive tree-building process.

In devising a vectorized algorithmic variant of NUTS, we note the following:
\begin{itemize}
    \item[-] Typical implementations of NUTS bound the number of allowed
    recursions (by default we cap \v{max_tree_depth} to 10).
    \item[-] The overall computational cost is dominated by gradient
    evaluations.
    \item[-] The remaining computation is dominated by U-turn checking.
    Various checks require a history of samples (made finite as implied by the
    first point).
\end{itemize}

With these observations in mind, NUTS is parallelized by ``unrolling'' the
recursion and implemented using \v{tf.while_loop}.
The general recipe is:
\begin{itemize}
    \item[-] Precompute program dynamism (e.g., recursion), noting read/write
    accesses.
    \item[-] Inspect the computation stack, noting repeat operations and
    requisite intermediate state.
    \item[-] Preallocate requisite memory (conceptually, the recursion stack)
    based on this analysis.
    \item[-] Use conditional operations \v{tf.where}, \v{tf.scatter} to update
    stack state. (With care taken to ensure constant computation across batch
    elements.)
\end{itemize}

For NUTS specifically, we note that naive data access could be implemented by
saving the entire trajectory of size $O(2^\text{\v{max_tree_depth}})$.
However, the pattern of access to state/momentum pairs is such that we need
only $O(\v{max_tree_depth})$ memory, i.e., exponentially less.\footnote{For
more details see
\href{https://github.com/tensorflow/probability/blob/master/discussion/technical_note_on_unrolled_nuts.md}{this technical note on the unrolled nuts implementation in TFP}.}

NUTS's candidate states are stacked during the tree doubling recursion;
naively this requires dynamic memory allocation. Following similar logic, we
observe it is possible to allocate a fixed-size memory buffer and update it in
situ. This ensures the leapfrog computation has a fixed memory access pattern
and further aids SIMD parallelism. The remaining bookkeeping necessary for a
NUTS transition (e.g., accumulating the energy difference) is straightforward.

\section{Second Contribution: A Python Callable Is All You Need}

While TFP provides sophisticated tools for model specification
\citep{dillon2017, piponi2020joint}, the MCMC API neither presumes
nor requires their use. This is possible since many MCMC techniques typically
only require evaluation of the (unnormalized) TLP function. For gradient-based
algorithms such as HMC, NUTS, and Langevin dynamics, automatic differentiation
capability is transparently provided by TensorFlow.


Bayesian applications of MCMC entail a joint generative model ``pinned'' at some observed data. In this case we recommend users specify the joint model distinct from the unnormalized. E.g.,
\begin{python}
jd = tfd.JointDistributionSequential([
  tfp.distributions.Uniform(low=0., high=1.),               # Prior
  lambda p: tfd.Independent(tfd.Bernoulli(probs=p[..., tf.newaxis]),
                            reinterpreted_batch_ndims=1),   # Likelihood
], validate_args=True)
unnormalized_posterior_log_prob = lambda pheads: jd([pheads, coin_flip_data])
\end{python}
(For more details, see \citep{piponi2020joint}.)

\section{Third Contribution: User-level Transition Kernel and Driver API}

Mechanistically, we can regard MCMC as a loop whose body consists of a single Markov transition. In TFP MCMC we reify these ideas as \v{TransitionKernel} and a ``driver.'' The \v{TransitionKernel} implements the chain mechanics, e.g., \v{tfp.mcmc.HamiltonianMonteCarlo}, \v{tfp.mcmc.NoUTurnSampler}, and composing kernels like \v{tfp.mcmc.MetropolisHastings}. \v{TransitionKernel}s implement \v{one_step} and \v{bootstrap_results}. The ``driver'' is parameterized by a \v{TransitionKernel} instance that relatively weakly specified by the API. Specialized drivers serve a variety of tasks, e.g., concatenate all samples, compute streaming statistics, interleave optimization (e.g., MCEM), or tune chain parameters.

Conceptually, TKs and drivers combine as follows,
\begin{python}
def driver(kernel, initial_state):
  [] = results
  side_results = kernel.bootstrap_results(initial_state)
  for _ in range(num_samples):
    x, side_results = kernel.one_step(results[-1], side_results)
    results += [x]
  return results
results = driver(SomeKernel(target_log_prob_callable), x0)
\end{python}

Per requirements of \v{tf.while_loop}, \v{TransitionKernel} ``\v{side_results}'' must be enumerable via \v{tf.nest}, itself supporting \v{tuple}, \v{list}, \v{dict}, and recursive combinations thereof. The containing structure and per-element \v{dtype}s of \v{bootstrap_results} must match that which is returned by \v{one_step}. Side-effects in \v{one_step} are not permitted; all state must be represented by either the chain state (\v{results[-1]}) or the \v{side_results}.

\subsection{TransitionKernel}

The \v{TransitionKernel} class encapsulates the computation of a single state transition in a Markov chain, as well as an initial ``bootstrap'' operation. These methods are called \v{one_step} and
\linebreak  
\v{bootstrap_results}, respectively.

The main work of the \v{TransitionKernel} is to generate a new chain state
from an existing one. We typically also need to keep track of additional
kernel-specific states, for example the Metropolis-Hastings
accept/reject outcome (for diagnostics) or the previous TLP
and its gradients (to avoid costly recomputation). The work of advancing the
chain state is done by the \v{one_step} method, which requires two arguments:
(1) the chain state, a \v{Tensor} or a \v{list}-like collection of
\v{Tensor}s, and (2) the kernel state, typically a (nested) \v{namedtuple} of
\v{Tensor}s. The output of \v{one_step} is a pair of new chain state and
kernel state. As noted above, the nature of TF loops demands these inputs
and outputs have identical structure. 

In order to start the chain, we need the initial chain state and kernel state. For end user, specifying the initial chain state is sufficient as the \v{bootstrap_results} method generate the suitable kernel-specific state from a chain state. For example, it may compute the value and gradients of
the TLP at initial chain positions. \v{bootstrap_results} method is typically called once as the code snippet above showed. 

The composable design of \v{one_step}---required by the TF loop API---means
we can naturally build up more complicated transition operations by nesting
\v{TransitionKernel}s. For example, MCMC samplers that include a
Metropolis-Hastings accept/reject step can be composed by nesting the
\v{MetropolisHastings} kernel with an ``uncalibrated'' transitional kernel: 

\begin{python}
randomwalk_mh = tfp.mcmc.MetropolisHastings(
    inner_kernel=tfp.mcmc.UncalibratedRandomWalk(
        target_log_prob_fn=target_log_prob_fn,
        new_state_fn=new_state_fn))
hmc = tfp.mcmc.MetropolisHastings(
    inner_kernel=tfp.mcmc.UncalibratedHamiltonianMonteCarlo(
        target_log_prob_fn=target_log_prob_fn,
        step_size=step_size))
\end{python}

Using a similar ``Matryoshka'' pattern, we can design \v{TransitionKernel}s
that internally call the \v{one_step} function of another
\v{TransitionKernel}, perform additional computation on the output chain state
and/or kernel state, and output the modified chain state and kernel state. For
example, we can perform parameter tuning during warmup/burnin by modifying the
inner kernel state (e.g. \v{tfp.mcmc.SimpleStepSizeAdaptation}).

An especially powerful example of this pattern is the
\v{TransformedTransitionKernel}, which employs \v{Bijector}s \cite{dillon2017}
to transparently apply smooth, volume-tracking reparameterizations of the chain
state space. Such reparameterizations are essential to algorithms like HMC,
which must operate in an unconstrained space, but can also be used for
preconditioning the state space to improve sampling efficiency \citep{hoffman2019neutra}.

\subsection{Driver}

At its simplest, a TFP MCMC driver is a Python function that iteratively
invokes a Markov transition operation and returns one or more intermediate
iteration values. To leverage TensorFlow's XLA compiler, low-level
optimizations, and hardware acceleration, TFP drivers rely on \v{tf.while_loop}
and derivatives \v{tf.scan}, \v{tf.map_fn}.

At present, TFP offers one driver, \v{sample_chain}, with 3 essential
features: burnin, sampling, and tracing. Samples produced during burnin are not
stored. Samples produced during sampling are, of course. Tracing allows the
user to obtain traces of data from the kernel state (for diagnostics).

Future plans include drivers for computing streaming expectations---useful,
e.g., when the number of state variables is so large that storing a full
trace becomes impractical---as well as ``multi-kernel'' drivers which allow
sequential composition of distinct \v{TransitionKernels}. The latter would, for
instance, enable more sophisticated hyperparameter adaptation phases before
burnin and sampling. 




%

\section{Discussion, Related Work, Future Work}

We described the MCMC module in TFP, including support for data parallelism,
a functional target log probability interface, and modular building blocks
for constructing performant MCMC algorithms runnable on modern hardware.

To the best of our knowledge, TFP MCMC is the first \emph{general} purpose 
library for \emph{constructing} MCMC algorithms, and making no strong demands on
the user's choice of model specification framework, other than the choice of
numerical library (TensorFlow) \footnote{Work is actively underway to enable
TFP to be used seamlessly with alternate numerical backends, currently NumPy
and JAX}. Some examples of more specialized libraries which do use data
parallelism include emcee \citep{Foreman_Mackey_2013}) and elfi
\citep{lintusaari2017elfi}). Additionally, NumPyro
\citep{phan2019iterativenuts} implements an iterative NUTS variant (similar in
spirit to TFP's, although the two developed independently), which is amenable
to JAX's program transformations, including JIT compilation (\v{@jit}) and
vectorizaiton via \v{@vmap}.

The availability of massively multi-chain MCMC provides new opportunities. For
example, running many parallel chains may enable adaptive MCMC techniques to
achieve faster convergence and lower bias, or allow for low-variance estimates
with relatively short chains. Separately, one typically computes convergence
criteria like potential scale reduction (R-hat
\citep{vehtari2019ranknormalization}) and effective sample size
\citep{geyer2011introduction, gelman2013bayesian} to monitor the convergence
of the chains. But the estimators that underlie these diagnostics
require relatively long chains, which blunts the advantage of running many
chains. Further research is needed to derive new tools and workflows that fully
exploit the potential of many-chain MCMC.

\section{Acknowledgements}
We would like to acknowledge contributions to the TFP library from Karl
Millar, Alexey Radul, Brian Patton, Srinivas Vasudevan, Danielle Kaminski,
Jasper Snoek, Sebastian Nowozin, and the TFP Team. We also thank Alexey Radul
and Ashok Popat for their review of this work.


\begin{thebibliography}{}

\bibitem[Abadi et~al., 2016]{abadi2016tensorflow}
Abadi, M., Agarwal, A., Barham, P., Brevdo, E., Chen, Z., Citro, C., Corrado,
  G.~S., Davis, A., Dean, J., Devin, M., et~al. (2016).
\newblock Tensorflow: Large-scale machine learning on heterogeneous distributed
  systems.
\newblock {\em arXiv preprint arXiv:1603.04467}.

\bibitem[Bingham et~al., 2019]{bingham2019pyro}
Bingham, E., Chen, J.~P., Jankowiak, M., Obermeyer, F., Pradhan, N.,
  Karaletsos, T., Singh, R., Szerlip, P., Horsfall, P., and Goodman, N.~D.
  (2019).
\newblock Pyro: Deep universal probabilistic programming.
\newblock {\em The Journal of Machine Learning Research}, 20(1):973--978.

\bibitem[Brooks et~al., 2011]{brooks2011handbook}
Brooks, S., Gelman, A., Jones, G., and Meng, X. (2011).
\newblock {\em Handbook of Markov Chain Monte Carlo}.
\newblock Chapman \& Hall/CRC Handbooks of Modern Statistical Methods. CRC
  Press.

\bibitem[Carpenter et~al., 2017]{carpenter2017stan}
Carpenter, B., Gelman, A., Hoffman, M.~D., Lee, D., Goodrich, B., Betancourt,
  M., Brubaker, M., Guo, J., Li, P., and Riddell, A. (2017).
\newblock Stan: A probabilistic programming language.
\newblock {\em Journal of statistical software}, 76(1).

\bibitem[Dillon et~al., 2017]{dillon2017}
Dillon, J.~V., Langmore, I., Tran, D., Brevdo, E., Vasudevan, S., Moore, D.,
  Patton, B., Alemi, A., Hoffman, M., and Saurous, R.~A. (2017).
\newblock Tensorflow distributions.
\newblock {\em arXiv preprint arXiv:1711.10604}.

\bibitem[Foreman-Mackey et~al., 2013]{Foreman_Mackey_2013}
Foreman-Mackey, D., Hogg, D.~W., Lang, D., and Goodman, J. (2013).
\newblock emcee: The mcmc hammer.
\newblock {\em Publications of the Astronomical Society of the Pacific},
  125(925):306–312.

\bibitem[Gelman et~al., 2013]{gelman2013bayesian}
Gelman, A., Carlin, J., Stern, H., Dunson, D., Vehtari, A., and Rubin, D.
  (2013).
\newblock {\em Bayesian Data Analysis}.
\newblock Chapman \& Hall/CRC Texts in Statistical Science. CRC Press.

\bibitem[Geyer, 2011]{geyer2011introduction}
Geyer, C.~J. (2011).
\newblock Introduction to markov chain monte carlo.
\newblock In {\em Handbook of Markov Chain Monte Carlo}, pages 139--188.
  Chapman and Hall/CRC.

\bibitem[Hastings, 1970]{hastings1970monte}
Hastings, W.~K. (1970).
\newblock Monte carlo sampling methods using markov chains and their
  applications.

\bibitem[Hoffman et~al., 2019]{hoffman2019neutra}
Hoffman, M., Sountsov, P., Dillon, J.~V., Langmore, I., Tran, D., and
  Vasudevan, S. (2019).
\newblock Neutra-lizing bad geometry in hamiltonian monte carlo using neural
  transport.
\newblock {\em arXiv preprint arXiv:1903.03704}.

\bibitem[Hoffman and Gelman, 2014]{hoffman2014no}
Hoffman, M.~D. and Gelman, A. (2014).
\newblock The no-u-turn sampler: adaptively setting path lengths in hamiltonian
  monte carlo.
\newblock {\em Journal of Machine Learning Research}, 15(1):1593--1623.

\bibitem[Lintusaari et~al., 2017]{lintusaari2017elfi}
Lintusaari, J., Vuollekoski, H., Kangasrääsiö, A., Skytén, K.,
  Järvenpää, M., Marttinen, P., Gutmann, M.~U., Vehtari, A., Corander, J.,
  and Kaski, S. (2017).
\newblock Elfi: Engine for likelihood-free inference.

\bibitem[Neal, 1993]{neal1993probabilistic}
Neal, R.~M. (1993).
\newblock {\em Probabilistic inference using Markov chain Monte Carlo methods}.
\newblock Department of Computer Science, University of Toronto Toronto,
  Ontario, Canada.

\bibitem[Neal, 2011]{neal2011mcmc}
Neal, R.~M. (2011).
\newblock Mcmc using hamiltonian dynamics.
\newblock In {\em Handbook of Markov Chain Monte Carlo}, pages 139--188.
  Chapman and Hall/CRC.

\bibitem[Phan and Pradhan, 2019]{phan2019iterativenuts}
Phan, D. and Pradhan, N. (2019).
\newblock Iterative {NUTS}.
\newblock \url{https://github.com/pyro-ppl/numpyro/wiki/Iterative-NUTS}.

\bibitem[Piponi et~al., 2020]{piponi2020joint}
Piponi, D., Moore, D., and Dillon, J.~V. (2020).
\newblock Joint distributions for tensorflow probability.

\bibitem[Robert and Casella, 2013]{robert2013monte}
Robert, C. and Casella, G. (2013).
\newblock {\em Monte Carlo Statistical Methods}.
\newblock Springer Texts in Statistics. Springer New York.

\bibitem[Salvatier et~al., 2016]{salvatier2016probabilistic}
Salvatier, J., Wiecki, T.~V., and Fonnesbeck, C. (2016).
\newblock Probabilistic programming in python using pymc3.
\newblock {\em PeerJ Computer Science}, 2:e55.

\bibitem[Ter~Braak, 2006]{ter2006markov}
Ter~Braak, C.~J. (2006).
\newblock A markov chain monte carlo version of the genetic algorithm
  differential evolution: easy bayesian computing for real parameter spaces.
\newblock {\em Statistics and Computing}, 16(3):239--249.

\bibitem[Vehtari et~al., 2019]{vehtari2019ranknormalization}
Vehtari, A., Gelman, A., Simpson, D., Carpenter, B., and Bürkner, P.-C.
  (2019).
\newblock Rank-normalization, folding, and localization: An improved
  $\widehat{R}$ for assessing convergence of mcmc.

\end{thebibliography}
\end{document}